\documentclass{PoS}

\title{Renormalized Polyakov loop in the Fixed Scale Approach}

\ShortTitle{Renormalized Polyakov loop in the Fixed Scale Approach}

\author{\speaker{Rajiv V. Gavai}%
\\
        Department of Theoretical Physics, Tata Institute of Fundamental Research Homi Bhabha Road, Mumbai 400005, India.\\
        E-mail: \email{gavai@tifr.res.in}}

\abstract{I compute Polyakov loop, the deconfinement order parameter, for
$SU(2)$ lattice gauge theory using the fixed scale approach for several
different scales and show how to obtain a renormalized physical order
parameter.  The generalization to other gauge theories, including quenched
or full QCD, is straightforward.  }

\FullConference{The XXVIII International Symposium on Lattice Filed Theory\\
		 June 14-19,2010\\
		 Villasimius, Sardinia Italy}

\begin{document}

\section{Introduction}

Phase transitions in various spin models have served as wonderful examples
for the study of quark-hadron transition in quantum chromodynamics (QCD)
and the related $SU(N_c)$ gauge theories, where $N_c$ is the number of
colours.  While average magnetization serves as the order parameter in the
former case, the Polyakov loop, $L$, defined as the product of the timelike
gauge links at a given site, is the order parameter for the deconfinement
transition \cite{McL}.  On an Euclidean $N_\sigma^3 \times N_\tau$ lattice
$L(\vec x)$ is defined at a site $\vec x$ as
\begin{equation}
L(\vec x) = \frac {1}{N_c}~{\rm Tr}~\Pi^{N_\tau}_{x_0=1}~U^4(\vec x, x_0)~,~
\end{equation}
where $U^\mu(x)$ are the gauge variables associated with the directed links
in the $\mu$th direction, $\mu=1$,4.  As in the spin models again, it is
convenient to define its average over the spatial volume, $\bar L =
\sum_{\vec x} L(\vec x) / N^3_\sigma$.  $\langle |\bar L| \rangle$ was used
to establish a second order deconfinement transition in numerical
simulations of the $SU(2)$ pure gauge theory.  Since then it has been used
for similar studies of the deconfinement phase transitions for a variety of
$N_c$ \cite{Cel}, for establishing the universality \cite{Gav} of the
continuum limit, as well as for theories with dynamical quarks \cite{DeT}.
Further, the predicted universality\cite{SvY} of critical indices has also
been numerically verified \cite{CrInd}.  Indeed, one hopes to be 
able to construct effective actions \cite{Pisa} for $L$ in a Wilsonian RG 
approach. These will be similar to the spin models in the same universality 
class but with possibly additional interaction terms.  A large number of 
models of quark-hadron transitions use the Polyakov loop as the order 
parameter for the deconfinement transition as well.

An order parameter should be physical, i.e., independent of the the lattice
size.  This is indeed so for spin models for sufficiently large lattices.
For $SU(N_c)$ gauge theories, this requirement means in addition
independence from the lattice spacing $a$ in the continuum limit.
Furthermore, it must be so in {\em both} the phases it seeks to
distinguish.  As is the case for any bare Wilson loop, the Polyakov loop,
needs to be renormalized for this to be true.  Since the bare Polyakov loop
is further known to decrease progressively with $N_\tau$, suggesting it to
be zero in the continuum limit in the high temperature phase, renormalized
$L$ is even more desirable to have.

\section{Results}

The physical interpretation of the order parameter as a measure of the free
energy of a single quark, $\langle \bar L (T) \rangle = \exp ( -F_Q(T)/T) $
provides a  straightforward clue for renormalization.  Since many years
various attempts to remove the divergent contribution in the single quark
free  in the continuum limit have been made.  These include computations
employing  lattice perturbation theory \cite{Hel},  use of the heavy
quark-antiquark free energy \cite{olaf1}, fits to $\langle \bar L \rangle$
on $N_\tau$-grids \cite{Dumi} and an iterative direct renormalization
procedure \cite{olaf2} for $\langle \bar L \rangle$ among others.

Here I advocate \cite{rvg10} another, perhaps better, method to define
renormalized $\langle \bar L \rangle$.   Let me elaborate why this maybe
so. The definition of Ref. \cite{olaf1} needs heavy quark potential at
short distances.  Lattice artifacts are at their worse when one is at such
short distances, with maximal violation of the rotational invariance.
Finite volume of the lattice also enters in defining the large distance
between the heavy quarks, or Polyakov loops.  Similarly the iterative
procedure used in Ref. \cite{olaf2} to obtain the renormalization constants
needs large lattices in both spatial and temporal directions.  Physically
perhaps an undesirable aspect of the definition of Ref. \cite{olaf2} is
that it works only on the plasma side, i.e., for $T \ge T_c$, where $T_c$
is the position of the peak in the Polyakov loop susceptibility.  The
definition \cite{olaf1} has so far been employed only in the $T \ge T_c$
for pure gauge theories for which $L$ is an order parameter.  It would
clearly be nice if the renormalization procedure is applicable to the
usually employed $\langle |\bar L| \rangle$, which is used as an order
parameter  on finite volumes.

I obtain a renormalized Polyakov loop which is valid for both the phases
below and above $T_c$ \cite{rvg10}.  It can be defined in any spatial
volume, and it becomes the true order parameter in the infinite spatial
volume limit.  Of course, it is also physical, i.e., $N_\tau$-independent
on finite volumes as well.  Indeed, it seems to work rather well for a
range of temporal lattice sizes, including $N_\tau \ge 4$.  I use the fixed
scale approach \cite{whot} to do so. It was introduced to minimize the
computational costs for the zero temperature simulations needed to subtract
the vacuum contribution in thermodynamic quantities such as the pressure
and to isolate pure thermal effects in computation of $T_c$
\cite{whotlat08}.  Furthermore, its advantage is that all the simulations
stay on the line of constant physics in a straightforward way.  What I
argue is that it is indeed this advantage which also permits an easy
renormalization of the Polyakov loop.  Although these considerations are
general, and apply to any $SU(N)$ gauge theory as well as any quark
representation, I shall consider below the simplest case of the $SU(2)$
lattice gauge theory to illustrate how and why it works.

\begin{figure}
\begin{center}
\includegraphics[scale=0.85]{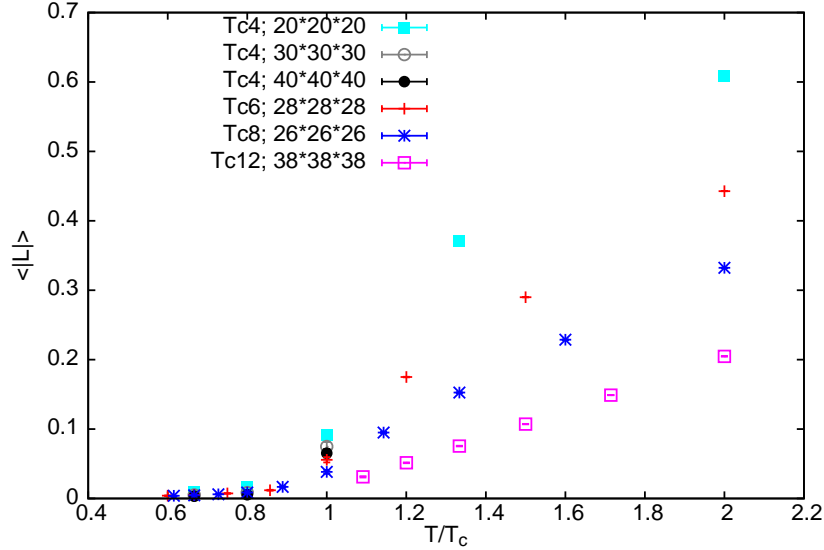}
\end{center}
\caption{The average Polyakov loop as a function of $T/T_c$ for four different 
scales.   The lattice sizes are as indicated in the key.  }
\label{fg.l68}\end{figure}

\begin{figure}
\begin{center}
\includegraphics[scale=0.85]{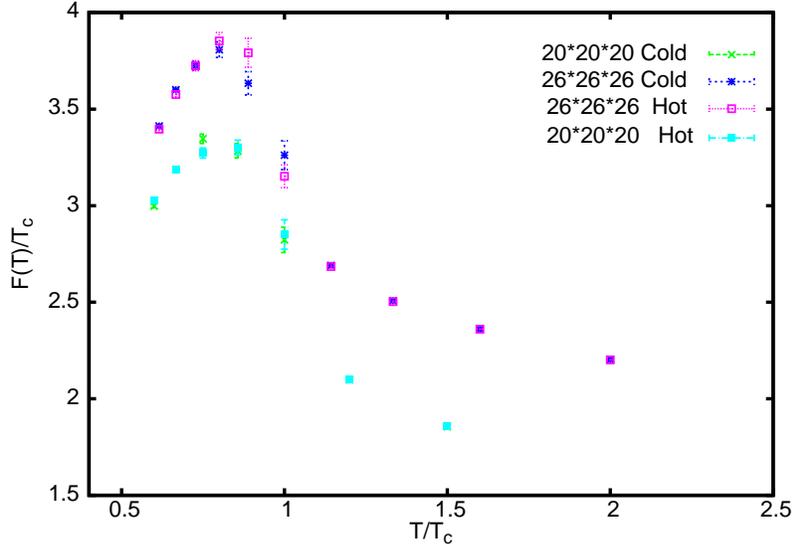}
\end{center}
\caption{The heavy quark free energy $F$ a function of $T/T_c$ for two 
different scales.   The lattice sizes are as indicated in the key.  }
\label{fg.bf68}\end{figure}

Recall that the temperature $T$ is varied in this approach by varying
$N_\tau$, holding the lattice spacing $a$, or equivalently the gauge coupling
$\beta = 2N_c/g^2$ fixed.  The single quark free energy $F_b(N_\tau, a)$ is 
then obtained from the $\bar L$ by the canonical relation, 
\begin{equation}
\ln \langle | \bar L | \rangle = -  a N_\tau F_b(N_\tau,a)~.
\label{eq.fe}
\end{equation}
The subscript $b$ reminds us that one obtains the bare free energy this
way.   If the chosen coupling is $\beta_c$, corresponding to the position
of the peak of the $|L|$-susceptibility in the usual fixed $N_\tau$
approach, and it lies in the scaling region, then the physical
deconfinement temperature $T_c = 1/N_{\tau,c} a_c$, and $T/T_c$ =
$N_{\tau,c}/N_\tau$ in the fixed scale approach, with the free energy given
by $F_b(T/T_c, a_c)$.  Writing it as a sum of a divergent and a regular
contribution, one has $a_c F_b(T/T_c, a_c) = a_c F(T/T_c,a_c) - a_c
A(a_c)$, where $A$ is the divergent free energy in physical units.
Clearly, the  divergent contribution will be {\em same} at all temperatures
in the fixed scale approach since it depends only on $a_c$.   

Since $\beta_c$, or $a_c$, is known precisely for the Wilson action of the
$SU(2)$ theory for many different $N_\tau$, I chose  four different scales
labelled $T_{c4}$, $T_{c6}$, $T_{c8}$, $T_{c12}$ corresponding to the known
transition couplings on $N_\tau = 4$, 6 \cite{Eng} and 8, 12 \cite{Vel}
respectively: $\beta_{c1} = 2.2991$, $\beta_{c2} = 2.4265$, $\beta_{c3} =
2.5104$, and $\beta_{c4} = 2.6355$.  Note that $T/T_c$ is given simply by
$n/N_\tau$ with $n= 4$, 6, 8 and 12 respectively.  Employing then $N_\tau =
3$ to 12, I varied the temperature in the range 2 $\ge T/T_c \ge 0.6$.
Note that fixed scale $a_c$ leads to a constant spatial volume in
physical units in each case in contrast to the usual fixed $N_\tau$
approach where the spatial volume varies with $T$.  I used a variety of
spatial lattice sizes. 
 
Figure \ref{fg.l68} shows the results for the thermal expectation value of
$\bar L $ as a function of the temperature in the units of $T_c$. In most
cases, I used both a random and an ordered start.  The errors are corrected
for autocorrelations.  The agreement in the data for the two starts suggest
the statistics of 200K iterations to be sufficient.   As expected, the four
different scales, $T_{c4}$ , $T_{c6}$ , $T_{c8}$ ,  and $T_{c12}$ lead to
four different curves for the order parameter.  One also sees the known
feature of $\langle \bar L  \rangle \to 0$ as $a_c \to 0$ even in the
deconfined region.  Figure \ref{fg.bf68} displays the behaviour of the bare
free energy for just two scales, obtained by using the eq.(\ref{eq.fe}).
The stars are for the scale $T_{c8}$ while the squares are for $T_{c6}$
corresponding to the higher lattice spacing of the two.  The figure
reinforces the expectation of the effect of the divergent free energy,
since the free energy increases with the decrease in the lattice cut-off $a_c$.

\begin{figure}
\begin{center}
\includegraphics[scale=0.8]{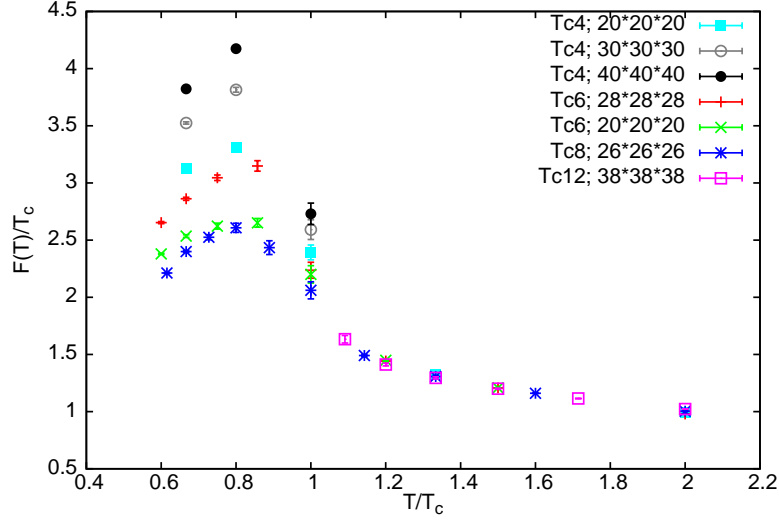}
\end{center}
\caption{ Heavy quark free energy $F$ as a function of $T/T_c$ 
with a constant shift, as explained in the text. }
\label{fg.f68}\end{figure}

\begin{figure}
\begin{center}
\includegraphics[scale=0.8]{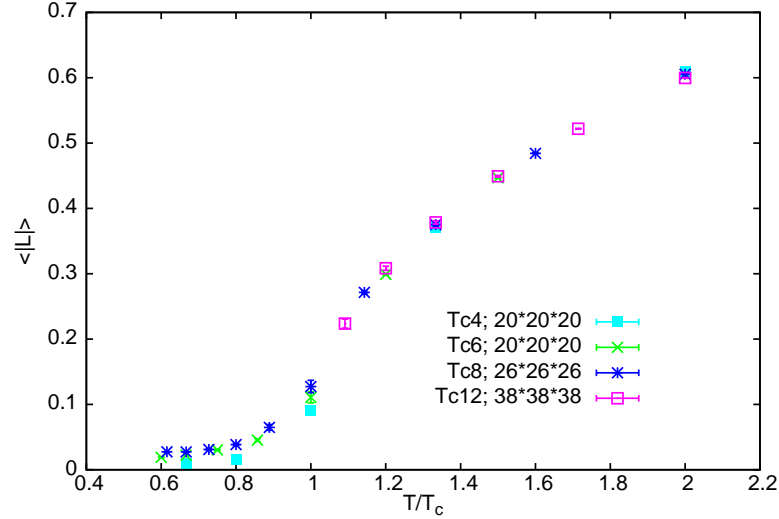}
\end{center}
\caption{Renormalized Polyakov loop versus $T/T_c$ using the shifted free energy
of the upper figure, as explained in the text.}
\label{fg.renl}\end{figure}
Any two different scales, $a_{c1}$ and $a_{c2}$ have their respective
divergent contributions, $a_{c1} A(a_{c1})$ and $a_{c2} A(a_{c2})$.
Multiplying eq.(\ref{eq.fe}) by $N_j$, for $j=1$
and 2 corresponding to the critical $N_\tau$ for the scale choices
above, i.e, 6 and 8, one obtains
\begin{equation}
\frac{T}{T_c}  \ln \langle | \bar L | \rangle = -  \frac
{F_b(T/T_c, a_{cj})}{T_c}~,~
\label{eq.fe1}
\end{equation}
where $F_b(T/T_c, a_{cj})/T_c = F(T/T_c,a_{cj})/T_c -  A(a_{cj})/T_c$.
Thus the free energies at the same temperatures but two different scales
are related by a mere constant, $[A(a_{c1})-A(a_{c2})]/T_c$.
For the four scales considered here, this implies 3 such constants.
Figure \ref{fg.f68} shows the results for the free energy 
with three constant shifts in the free energy determined by demanding
coincidence at the 
highest $T =2 T_c$.   A universal curve for the free energy seems to
result as a result for a wide range of $T > T_c$.  The results for
the low temperature phase are seen to be volume dependent, as expected.
In the infinite volume limit, the free energy should increase to infinity in
the confined phase whereas it should essentially remain constant in the
deconfined phase. Such an expectation is indeed borne out by the results
in the Figure \ref{fg.f68}.  For the same physical volume, the free energy
appears to be $a$-independent in the $T < T_c$ phase as well, as seen by
comparing the crosses and the stars.

Finally, it should now be clear how one can obtain a universal curve for
the order parameter from the universal free energy curve.  The $\langle |
\bar L | \rangle $ corresponding to scale $\beta_{c2}$ should simply be
multiplied by the factor $\exp(N_\tau[A(a_{c1})/T_c -A(a_{c2})/T_c])$ and
then the date will lie on a universal curve.  This is exhibited in Figure
\ref{fg.renl} for {\em all} the four scales.  It is worth noting that the
same universal order parameter results in {\em both} below and above $T_c$
by fixing only three constants for the four scales exhibited.  The entire
low and high temperature region of the order parameter is uniquely fixed,
and appears to be universal.

\begin{figure}
\begin{center}
\includegraphics[scale=0.825]{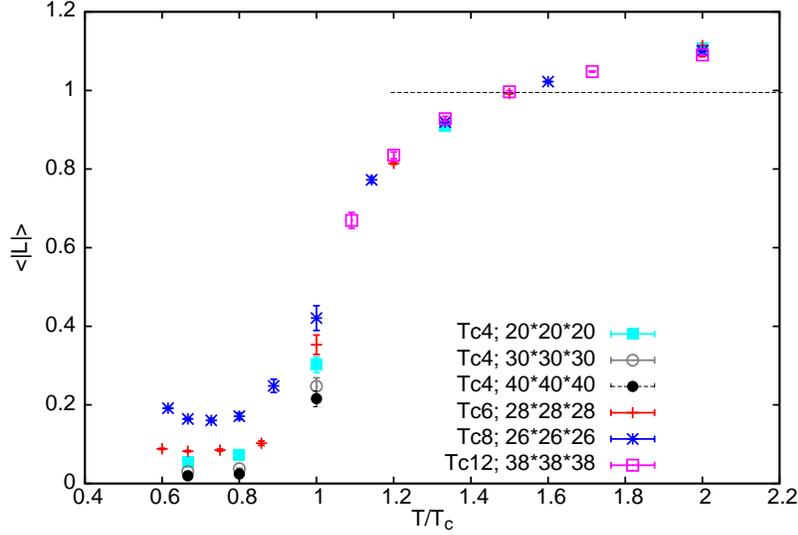}
\end{center}
\caption{Renormalized Polyakov loop versus $T/T_c$ after subtracting the
would-be divergent contribution.}
\label{fg.perl}\end{figure}

From the  Figure \ref{fg.renl}, it appears as though the approach
of  $\langle | \bar L | \rangle $ to unity is slow and from below.
It is, however, known since long \cite{GaJe} that perturbation theory
predicts $L \to 1$ from {\em above} at very large $T$ : $ L = 1 + C_3 g^3 +
{\cal O}(g^4)$, where $c_3(N_c)> 0$ is a constant.  The solution to this
apparent paradox can be traced to the usual fact that a renormalized
quantity depends on the scale chosen to define the scheme for
renormalization.  In my case, the inclusion of a constant free energy
$A(a_{c})/T_c$  for the chosen scale $a_c$ defines the choice.  The details
of the shape of the physical order parameter are therefore scale-dependent
in the plasma phase but it is universal none the less once a choice is
made.  Moreover, any further change of scale leads to a computable change
in the shape.  Indeed, in order to mimic the perturbative renormalization
scheme I estimated the point-divergent contribution.  At the highest
temperature 2$T_c$, I fitted the results for the four scales to $-\ln
\langle | \bar L_j | \rangle =  F(2T_c)/2T_c + B \cdot N_{\tau j}/2$.
Having thus determined the coefficient $B$ of the would-be divergent
contribution at the scale $T_{c4}$,  I eliminated the $B$-dependent
contribution at that scale.  The renormalized $\bar L$ at the other three
scales were related to it by the same shifts as before.  Figure
\ref{fg.perl} displays the resultant $\langle | \bar L | \rangle $.  It
crosses unity  at about 1.5 $T_c$. Since all the multiplying factors tend
to unity at large $T$, the approach of this $\bar L$ to unity is from above
at large $T$.  Note also that large spatial volumes, aspect ratio of $ \sim
10$, are needed for this $L \simeq 0$ in the low $T$ phase.

\section{Summary}

In conclusion, I showed that the fixed scale approach leads to a natural
definition of a physical, $N_\tau$-independent, order parameter which is
defined in both the confined and the deconfined phases. The definition
itself does not depend on any lattice artifacts or the lattice size in the
deconfined phase , and works very well for even coarse lattices ($ a \le
1/4T_c$).  Moreover, it displays the expected behaviour in the confined
phase as the physical volume is increased, suggesting that the so
determined physical free energy of a single quark in the confined phase,
$F$, goes to infinity in the infinite volume limit.  Eliminating the
point-divergent contribution leads to a high temperature behaviour
consistent with perturbation theory.  It is straightforward to generalize
this idea to $SU(N_c)$ gauge theories and QCD as well as to sources in
higher representations.

\end{document}